\documentclass{article}
\usepackage[english]{babel}

\usepackage[letterpaper,top=2cm,bottom=2cm,left=3cm,right=3cm,marginparwidth=1.75cm]{geometry}

\usepackage{amsmath}
\usepackage{natbib}
\usepackage{graphicx}
\usepackage{booktabs}
\usepackage[colorlinks=true, allcolors=blue]{hyperref}
\usepackage{authblk}

\usepackage{placeins} % for FloatBarrier

\title{Multi-label classification of open-ended questions with BERT}
\author[1]{Matthias Schonlau\thanks{Email: schonlau@uwaterloo.ca}}
\affil[1]{Department of Statistics and Actuarial Science, University of Waterloo}
\author[2]{Julia Weiß}
\author[2]{Jan Marquardt}
\affil[2]{GESIS, Leibniz Institute for the Social Sciences, Mannheim}
\begin{document}
\maketitle

\begin{abstract}
Open-ended questions in surveys are valuable because they do not constrain the respondent's answer, thereby avoiding biases. However, answers to open-ended questions are text data which are harder to analyze.  
Traditionally, answers were manually classified  as specified in the coding manual. In the last 10 years, researchers have tried to automate coding. Most of the effort has gone into the easier problem of single label prediction, where answers are classified into a single code. However, open-ends that require multi-label classification, i.e., that are assigned multiple codes, occur frequently. 
In social science surveys, such open-ends are also frequently mildly multi-label. In mildly multi-label classifications, the average number of labels per answer text is relatively low (e.g. $<1.5$). For example, the data set we analyze asks ``What do you think is the most important political problem in Germany at the moment?" Even though the question asks for a  single problem, some answers contain multiple problems. Of course, the average number of problems (or labels) per answer is still low.

This paper focuses on multi-label classification of text answers to open-ended survey questions in social science surveys.  We evaluate the performance of the transformer-based architecture BERT for the German language in comparison to traditional multi-label algorithms (Binary Relevance, Label Powerset, ECC) in a German social science survey, the GLES Panel (N=17,584, 55 labels). Because our data set requires at least one label per text answer, we also propose a modification in case the classification methods fail to predict any labels. We evaluate the algorithms on 0/1 loss: zero loss occurs only when all labels are predicted correctly; a mistake on one label incurs the full loss (1). This loss  corresponds to the reality of manual text classification: you code the whole text answer with all labels, even if only a single suspicious label requires review. 

We find that classification with BERT (forcing at least one label) has the smallest 0/1 loss (13.1\%) among methods considered (18.9\%-21.6\%). As expected, it is much easier to correctly predict answer texts that correspond to a single label (7.1\% loss) than those that correspond to multiple labels ($\sim$50\% loss). Because BERT predicts zero labels for only 1.5\% of the answers, forcing at least one label, while  successful and recommended, ultimately does not lower the 0/1 loss by much.

Our work has important implications for social scientists: 
1) We have shown multi-label classification with BERT works in the German language for open-ends. 
2) For mildly multi-label classification tasks, the loss now appears small enough to allow for fully automatic classification. Previously, the loss was more substantial, usually requiring semi-automatic approaches.
3) Multi-label classification with BERT requires only a single model. The leading competitor, ECC, is an iterative approach that iterates through individual single label predictions.

\end{abstract}
%%%%%%%%%%%%%%%%%%%%%%%%%%%%%%%%%%%%%%%%%%%%%%%%%%%%%%%%%%%%%%%%%%
\section{Introduction}
Open-ended questions are useful because they avoid bias that may arise by constraining respondents' answer choices. However,  the resulting text answers are inconvenient to analyze. As a consequence,  open-ended questions may be underutilized in survey research.  When open-ended questions are indispensable, they are typically manually classified following a pre-existing coding scheme.

In the last 10 years, research has moved towards classifying text answers to open-ended questions  automatically \citep{schonlau2016semi,schierholz21occupation,he2020double_multiple}. 
Usually, some text answers (e.g. 300-500) are coded manually and form the training data. Then, following the bag-of-words approach, ngram variables \citep{schonlau17ngram} are computed. Ngram variables are counts of how often a word or a word sequence appears in the answer text. Finally, using supervised learning, a learner is trained on the training data and used to classify the answers texts. 
Because accuracy is often not high enough for research purposes, semi-automatic algorithms have been proposed \citep{schonlau2016semi} that classify easy-to-classify answers automatically and hard-to-classify answer manually. 

Most of the effort has gone into studying classification of text answers of open-ends into exactly one class. This is called single label classification.  Text answers can also be tagged with  multiple labels; for example, if the respondent mentions more than one theme in their answer.
This is analogous to the check-all-that-apply questions where multiple answers are also possible.
Such multi-label classification is much harder: the classifier has to classify multiple labels (rather than just one label) correctly  and at the same time the classifier has to figure out how many labels are appropriate for a given text answer.

Even though many open-ended questions are multi-label, relatively little research has investigated multi-label open-ended questions in the social sciences. In particular, a comparison of the performance of a transformer-based architecture such as BERT to standard multi-label algorithms is lacking.  
Research using BERT to analyze social science open-ended questions in a language other than English is also lacking. 

The outline of the remainder of this paper is as follows. 
Section~\ref{s:background} reviews multi-label algorithms, BERT, and multi-label classification for open-ended questions in social science surveys. 
Section~\ref{s:methods} introduces the German data set used in this paper and the experimental setup. 
Section~\ref{s:results} gives the results of BERT vs. the multi-label classification algorithms. 
Section~\ref{s:discussion} concludes with a discussion.

\section{Background\label{s:background}}

\subsection{Multi-label classification algorithms\label{s:multi-labelalg}}
The easiest and possibly most popular approach to classifying multi-label data is binary relevance (BR) \citep{tsoumakas2007}.  BR treats the multi-label problem as L separate single-label regression problems where L is the number of labels. This approach is easy to implement but ignores the correlation among the labels. 

 The Label Powerset (LP) \citep{tsoumakas2007}  method transforms the multi-label problem to a single-label problem: it treats each labelset (combination of labels) that occurs in the training data as a single label.  For example, for 10 labels, there are 10 choose 2 possible label combinations. Even though many of these labelsets will not occur in the training data, the number of labelsets will likely still be large.  
 LP is a single model and the model accounts for correlation among labels by only allowing for labelsets that were seen in the training data. 
 While LP does take into account correlations among labels, the large number of classes can make this method useless in practice. A possible exception are mildly multi-label data where the average number of labels is small, creating fewer label combinations.  
 
Classifier Chains (CC) is an iterative method that incorporates correlations by including indicator variables of other labels as additional x-variables \citep{read2011classifierchains}.  As in BR, labels are predicted one at a time: The first label is predicted as usual.  The regression for the second label includes the predicted first label as an additional x-variable. The regression for the third label includes two additional x-variables with the predicted first and second labels, and so forth.  To reduce the dependence on the  order of labels, the authors proposed an ensemble version of Classifier Chains, ECC. 
ECC runs CC repeatedly on bootstrap samples using random label order. 
 Because ECC is superior to CC, typically only ECC results are reported. 
ECC tends to have high predictive performance \citep{madjarov2012extensive,schonlau2021all-that-apply}.

Additional methods to multi-label classification are described, for example, in  \citet{tsoumakas2007} and \citet{zhang2013review}.

\subsection{BERT}
Bidirectional Encoder Representations from Transformers (BERT) \citep{devlin2019bert} is a language  model  based
on the Transformer architecture \citep{vaswani2017attention}.  
BERT is able to consider both the left and right contexts of a masked (or hidden) word, which is referred to as bi-directional attention.
BERT converts input words into numerical vectors. During 
pre-training, the numerical representation of words 
as well as more than 1 million coefficients are optimized to predict masked (hidden) words as well as possible. (There is also a second task, next sentence prediction, which is of lessor importance).
The purpose of pre-training is to learn a language model -- i.e. predicting masked words well -- on a vast amount of unrelated text data. Fine tuning then optimizes a criterion related to classification using the specific data of interest. For single label classification, BERT has one output variable (y-variable) per class. Suitable loss and activation functions ensure that the estimated probabilities for each class sum to 1. The class with the largest probability is predicted.
BERT can also be used for multi-label classification by specifying different loss and activation functions for the output variables.
For multi-label classification, BERT also has one output variable per class. However, the estimated probabilities no longer have to sum to 1. The set of predicted labels are all labels for which the corresponding probability is larger than a threshold probability, usually $0.5$.

In summary, multi-class and multi-label classification are surprisingly similar: they use the same language model, and they have the same number of output variables, one per class. They just differ in the constraints on the output probabilities and the loss function. Because the multi-label predictions arise from a single model rather than from multiple independent models, no further effort is required to account for correlations among the labels.  

\subsection{Multi-label classification for open-ended questions in social science surveys}
Multi-label classification for open-ended questions has been relatively understudied. Because multi-label algorithms are difficult problems with low subset accuracy (subset accuracy and 0/1 loss are equivalent; they both require predicting {\em all} labels of a text answer correctly), \citet{gweon2020semi} proposed a semi-automatic algorithm for coding multi-label open-ended questions.
 The idea is to classify {\em automatically} when the algorithm is reasonably sure that the classification is correct, and {\em manually} otherwise. Such semi-automatic classification ensures overall high accuracy. 
 
When encoding text using ngram variables, multi-label classification algorithms (see Section~\ref{s:multi-labelalg}) for open-ended questions performed better than BR  using 0/1 loss \citep{schonlau2021all-that-apply}.  The performance appeared to be better the stronger labels were correlated with one another.

 \citet{meidinger2021open} provide 19 English language multi-label benchmark data sets and evaluate them using BERT and other transformer-based models. Surprisingly, they find that the baseline model, logistic regression (one vs. rest), beat BERT 3 of 10 times in terms of 0/1 loss.

 \citet{ford2023} used BERT to encode the text answers (rather than for classification). The encodings are then used as input to logistic regression; one logistic regression for each label. The logistic regressions are independent from one another, but they share the same x-variables. The text data are in English.

The average number of labels (also called {\em cardinality}) of multi-label classification tasks is one way to characterize the classification task. For the purpose of this paper, we  call a classification task {\em mildly multi-label} if it contains fewer than 1.5 labels per text answer on average. A substantial fraction of multi-label open-ended questions is only mildly multi-label. 
For example, if respondents are asked to supply only one reason, 
even if some respondents ignore this request, 
the data will have  a small average number of reasons or labels.
Elsewhere,  the average number of labels in three data sets was reported as  1.15, 1.44 and 2.77 labels \citep{schonlau2021all-that-apply}.  Among the data sets reported in  \citet{meidinger2021open}, 4 of 10 are only mildly multi-label.

%%%%%%%%%%%%%%%%%%%%%%%%%%%%%%%%%%%%%%%%%%%%%%%%%%%%%%%%%%%%%%%%%%
\section{Methods\label{s:methods}}
We compare multi-label BERT with several multi-label algorithms on a large data set.
We first describe this data set (\ref{s:data}),
then we give details about the architecture and tuning of the models (\ref{s:models}), 
we consider the special case where a model predicts zero labels (\ref{s:zerolabel}), and we discuss the evaluation criterion (\ref{s:criterion}). 
%-----------------------------------------------------------
\subsection{Data\label{s:data}}
We use data from wave 1 of the election campaign panel (2016) in  the  German Longitudinal Election Study (GLES)  \citep{glespanel}.
This Web survey asked the following question in German ``What do you think is the most important political problem in Germany at the moment? Please name just one problem."  
The survey was drawn from an access panel with quota for gender (2 categories), education (3 categories), and age (5 categories). 18,128 respondents participated. Of these, 97.0\% responded to the open-ended question, resulting in 17,584 text answers. Here, ``responded'' means the answer consisted of at least one character. 
 All non-empty answers were part of the prediction, but any nonsensical  answers were classified as ``no answer'' (see Table~\ref{t:top10} for examples).
%considered interviews (incl. 30 breakoffs before question): 18128 \\
%gave any open-ended answer: 98,49\% \\
%gave non-missing open-ended answer (-99, -98 excluded): 17129/ 18128 = 94,49\% }

The average answer length is 3.57 words. 61.8\% of the answers contain just one word; the third quartile is 3 words. The maximum word length is 333 words.

Each answer was manually coded into one or more of 55 different answer labels.
 The maximum number of labels is 5. 
The average number of labels is 1.17.
$12.6\%$ of the answers have more than one label.  
Table~\ref{t:top10} lists the 10 most common labels and their percentage (out of all labels).
\begin{table*}[tbhp]
\centering
\caption{Top 10 labels, percentages, and examples. The percentage refers to the total number of labels, not to the total number of answer texts. }
\label{t:top10}
\begin{tabular}{lrrl} \toprule
Label name & code &  \%  & Examples  \\ \midrule
immigration, asylum and integration & 3750  &  47.93  &``close borders'',``too many new immigrants''  \\
social justice & 3720  &  5.38 & ``inequality'', ``equal opportunities'' \\
values, political culture, social criticism & 2400  &  4.40 &  ``tendency to the right''  \\
radicalisation, extremism & 2411  &  3.35 &  ``fighting extremism", ``Islamic threat" \\
poverty & 3721  &  3.03 & ``unjust social system", ``social problems" \\
no answer &-99    &  2.84  & ``…'', ``xxx'' \\
demographic change and retirement & 3740  &  2.78 & ``social security payments" \\
political structures and processes & 2000  &  2.60 &  federalism, chancellor \\
terrorism & 3412  &  2.57 &  ``terror", ``attacks" \\
criticism of political parties (specific)& 2439  &  1.61 &  naming a political party \\
\bottomrule
\end{tabular}
\end{table*}
  In this data set, the immigration label (3750) accounts for almost half the labels. Only a couple of percent of the labels are ``no answer'', and even fewer are ``don't know'' or in the catch-all category ``other class'' (not among the top 10 labels).
When the unit of analysis is an individual  label, the interrater-reliability kappa is $\kappa=0.88$. 
%(For multiple labels each can be separately correct or incorrect.)
When the unit of analysis is the text answer, where the set of all labels have to be simultaneously correct, the interrater-reliability kappa is  $\kappa=0.82$. 
%kappa after Label Powerset transformation: 0.82 (agreement: 87\%) \\
%kappa long format binary: 0.90 (99.6\% agreement)}

%-----------------------------------------------------------
\subsection{Models\label{s:models}}
We evaluate several algorithms for multi-label predictions: multi-label BERT, binary relevance, Label Powerset, and Classifier Chains. Here, we give details about the algorithms used. 
The data were split at random into training data (60\%), evaluation data (20\%), and test data (20\%). We used the same split for all algorithms.

\subsubsection*{BERT}
Our language model is bert-base-german-cased \citep{germanbert}.
``base'' refers to 12 encoder layers stacked on one of top of the other, 12 attention heads and 768 hidden units. The model has about 110 million parameters. We access the model through Thilina Rajapakse's ``Simple Transformers'' Python package, which is built atop Pytorch.

We use 55 output variables corresponding to the 55 classes in our data set. Each output variable is an indicator variable.
 We performed a grid search on the number of epochs (5, 10, 20) and the learning rate (1e-3, 1e-4, 1e-5) on the evaluation data. The final run used 20 epochs  and a  learning rate of $lr=1e-5$. There were minimal difference between the loss function of the final run and a run with the default learning rate, $lr=1e-4$, and either 10 or 20 epochs.

\subsubsection*{Preprocessing}
Pre-processing applies to the multi-label algorithms that use the ngram variables. 
  We removed ``words'' that consisted only of a single letter.  In German, single-letter-words do not exist (unlike in English, e.g., ``a",``I").
Second, respondents wrote German umlauts inconsistently (ä vs ae, ü vs ue, ö vs oe, and ß vs ss). For this reason, we transformed all umlauts into their respective alternative form (ae, ue, oe, ss). 
Third, two German political parties have names that can be confused with a  corresponding regular German word.  The green party ``die Grünen" is also  a color and  ``die Linke" also means ``left''.  To disambiguate the political  party from the corresponding word, we used a regular expression to replace any mention of the party with  (\verb|die_grünen| and \verb|die_linke|) before creating ngram variables. We also lower-cased all words, removed punctuation, and removed all numbers. We did not remove stopwords to avoid accidentally removing an important word.

We then used the bag-of-words approach to turn the text into numerical variables. Specifically, we used the HanTa Lemmatizer \citep{wartena2019},  a simple approach to lemmatization based on Hidden Markov Models that is trained for the German language. The HanTa Lemmatizer created unigrams (counts of single words). Finally, we applied the  TF-IDF transformation. This transformation assigns more weight to   more unusual words and to larger counts.

\subsubsection*{SVM and Binary Relevance}
We used the support vector machines implementation in  the “scikit-learn” library for Python \citep{scikit-learn}. For binary relevance, we employed the ``one vs. rest'' classification approach using support vector machines as the base learner with the usual threshold, 0.5.  
For both a linear and a radial kernel, we performed a grid search to determine a suitable values for the hyperparameters $C$ and gamma (gamma applies only to the radial kernel) to minimize 0/1 loss. The x-variables are the TF-IDF variables. The models were trained on the training data and evaluated on the validation data. For binary relevance, the grid search determined a linear kernel with  $C=100$.

\subsubsection*{Ensemble Classifier Chains}
We created 10 Classifier Chains, using the “ClassifierChain” function from “sklearn.multioutput” with SVM as a base learner. We used the usual classification threshold of 0.5.  We used the same SVM parameters as for the Binary Relevance approach, i.e. a linear kernel with $C=100$.

\subsubsection*{Label Powerset}
There were 813 unique combinations of the 55 labels in the training data.
Label Powerset only requires one single-label classifier.
As before, we used a grid search to optimize the 0/1 loss on the validation data. For the Label Powerset method, the grid search determined a radial kernel with $C=100 $ and $gamma=0.5$. 

%-----------------------------------------------------------
\subsection{When zero labels are predicted\label{s:zerolabel}}
All answer texts in the  data set have at least one label. If the answer does not fit a subject matter class, 
it is given one of the labels  ``no answer'', ``don't know'' or ``other class'' as appropriate.
While the texts always have a label, the algorithms may predict zero labels. (The Label Powerset algorithm is an exception: because Label Powerset transforms the problem into a single-label prediction, it must predict at least one label.)
We know that zero-label predictions must be incorrect. Therefore, when the multi-label algorithm predicts zero labels, we substitute the label that single-label prediction predicts. In the table we refer to this method as ``BERT min 1 label''. The same approach could also be used for other algorithms, but for clarity of presentation we only apply it to the winning algorithm, BERT.

%-----------------------------------------------------------
\subsection{Evaluation Criterion\label{s:criterion}}
For single label prediction, there is only one way to compute loss or, equivalently, accuracy. In the presence of multiple labels, there are at least two loss functions in common use: 0/1 loss and Hamming loss. The 0/1 loss is 0 (no loss) when all labels are predicted correctly and 1 otherwise.  This is a very strict criterion, as even a single mistake among our 55 predicted labels would incur the full loss. Hamming loss computes the average number of labels predicted incorrectly. This loss function is more forgiving and may be useful when the average number of labels is moderately high or high. 

We have 55 labels but the average number of labels is 1.17. 
Because of the low average number of labels, the 0/1 loss 
criterion is appropriate here. 
The Hamming loss is extremely small here
as most of the 55 labels will always be correctly predicted as 0.

Other evaluation criteria popular in computer science papers include macro F, micro F, and LRAP. (LRAP takes into account where the true label ranks among the predicted labels.)  We argue that 0/1 loss is the most relevant criterion in the social sciences. If an answer text is suspected to be misclassified, a coder has to manually code the answer text. This is the same effort regardless of whether one or multiple labels are incorrect.

%%%%%%%%%%%%%%%%%%%%%%%%%%%%%%%%%%%%%%%%%%%%%%%%%%%%%%%%%%%%%%%%%%
\section{Results\label{s:results}}
In Section~\ref{s:mainresults}, we evaluate 0/1 loss for BERT, binary relevance, Label Powerset, and  Ensemble Classifier Chains (ECC). For BERT, we also evaluate zero-one loss after forcing at least one label. 
Section~\ref{s:bad} considers bad predictions and what we can learn from them. 
In light of BERT's particular strengths fro single label predictions, Section~\ref{s:semiautomatic} proposes a semiautomatic approach.
\subsection{Main results\label{s:mainresults}}
Table~\ref{t:zerooneloss} shows 0/1 loss by method overall and by the true number of labels.  BERT and BERT with at least one label have the smallest 0/1 loss by far. They misclassify just over 13\%  text responses based on 0/1 loss. As expected, Binary Relevance does worst. 
\begin{table*}[htbp]
\centering
\caption{0/1 loss by number of labels on the test data by method and true number of labels. The algorithms are sorted by increasing overall loss.}
\label{t:zerooneloss}
\begin{tabular}{lrrrrrr} \toprule
  &  Overall & 1 label & 2 labels &  3 labels & 4 labels & 5 labels  \\ \midrule
 BERT min 1 label  &
 0.1308 &
 0.0705 &
 0.4893 &
 0.8225 &
 0.9130 &
 1.0000\\
 BERT &
0.1347 &
 0.0749 &
 0.4893 &
 0.8225 &
 0.9130 &
 1.0000\\
 ECC  &
 0.1891&
 0.1315&
 0.5319&
 0.8387&
 0.9565&
 1.0000\\
Label Powerset &
 0.1959 &
 0.1063 &
 0.8116 &
 1.0000 &
 0.9565 & 
 1.0000   \\
Binary Relevance &
 0.2161 &
 0.1464 &
 0.6657 &
 0.9032 &
 1.0000 & 
 1.0000 \\
\bottomrule
\end{tabular}
\end{table*}

Looking at the loss by true number of labels, we find that the responses with multiple labels are far harder to predict than responses with a single label. For BERT, the 0/1 loss jumps from about 7\% for one label to about 49\% for 2 labels. In fact, the more labels there, the worse the 0/1 loss.

The distribution of the number of labels for the algorithms is shown in Table~\ref{t:label_distribution}.
\begin{table*}[thbp]
\centering
\caption{Distribution of the number of labels (\%) in the test data by prediction method }
\label{t:label_distribution}
\begin{tabular}{lrrrrrr} \toprule
  & 0 labels &  1 label & 2 labels &  3 labels & 4 labels & 5 labels  \\ \midrule
true  
& 0.0
&   88.0
&    9.4 
&    1.8
&    0.7 
&    0.3 \\
BERT 
&    1.5
&    89.0
&    8.1
&    1.1
&    0.2 
& 0.0 \\
ECC 
&   2.8
&   88.4
&   8.0
&   0.8
&   0.0
&   0.0
\\
Label Powerset
& 0.0
& 96.0
& 3.6
& 0.3
& 0.0
& 0.0  \\
Binary Relevance 
&8.7
&84.9
&5.9
&0.5
&0.0
&0.0  \\
\bottomrule
\end{tabular}
\end{table*}
As previously mentioned, the data set is only mildly 
multi-label:  88\% of the data are single label.
Compared to BERT, the other algorithms predict multiple labels less often.
Binary Relevance does particularly poorly, predicting zero labels 8.7\% of the time. While the performance could be improved by forcing at least one label, 
it is clear it would not do as well as BERT 
because BR also performs worse when the number of labels is larger (see Table~\ref{t:zerooneloss}).
Label Powerset cannot predict zero labels because the multi-label problem is transformed to a single label problem. 

When forcing BERT predictions to have at least one label (``BERT min 1 label''), the decrease in overall loss is small in absolute terms, about $0.4$\% (Table~\ref{t:zerooneloss}).  However, only $1.5$\% of BERT's predictions had zero labels (Table~\ref{t:label_distribution}).  Keeping in mind that these answers were difficult to classify for BERT, classifying $0.4$ of $1.5$ percentage points correctly is quite good.

\subsection{Bad predictions\label{s:bad}}
BERT assigned a small number of answers no label even though all answers had at least one label. The word count of these answers tends to be higher than that the average word count of the entire data set.  Answers predicted to have zero labels include, for example, ``ich bin in politischen dingen neutral!'' (``I am neutral in political things''), and the single word answers ``boeh'' (meaning unclear) and ``Menschlichkeit'' (``humanity'').  
The true labels of these answers was very diverse.

Next, we consider single-label answers that are also predicted to be single-label by BERT and ECC but with an incorrect label. ECC  often incorrectly assigned -99 (``no answer'') or the dominant (most frequent) category. When ECC incorrectly assigned ``no answer'', the answer often contained spelling errors.  During pre-processing we did not attempt to correct for spelling errors and this appears to have affected ECC much more than BERT.  

When both ECC and BERT predict incorrect labels, the correct codes are most often 
2400 (``values, political culture, social criticism"), 2000 (``political structures and processes") and 5000 (``other").  All three categories have in common that they cover a wide range of topics.  Table~\ref{t:2400} gives some examples when the true code is $2400$; Table~\ref{t:Appendix1} in the Appendix shows the corresponding original text answers in German.  Strikingly, ECC and BERT often predict the same false labels and  these false labels are often reasonable. 
For example, ``constant violations of our basic law and mrs. merkel's policy without alternatives"  was assigned to code 2441 "criticism of politicians, specific" by both ECC and BERT but was manually coded as the more general code 2400 (see Table~\ref{t:2400}). Code 2441 is also correct, since specific criticism was levelled at the then German Chancellor Angela Merkel. The difference in labeling is perhaps better described as an ambiguity in the coding manual between a general and a more specific code.

\begin{table*}[thbp]
\centering
\caption{Incorrectly predicted codes by ECC and BERT when the correct code 2400. Explanations of common codes are shown in Table~\ref{t:top10}.  Code  5000 is ``other''.}
\label{t:2400}
\begin{tabular}{lrrrrrr} \toprule
 Answer & ECC &  BERT   \\ \midrule
``they do not look after the german"  
& 3100
&   5000
 \\
``egoism"
 
&    -99
&    5000
 \\
``creating satisfaction"
&   5000
&   5000

\\
``unclear goals"
& -99
& 5000
 \\
``better explanation of politics" 
&2000
&2000
  \\
``too many non-voters"
&-99
&-98
  \\
``consistent implementation"
&-99
&2000
  \\
``uniformity"
&-99
&5000
  \\
``constant violations of our basic law and mrs. merkel's policy 
&2441
&2441
  \\
without alternatives" &2441 &2441 \\
``communication"
&5000
&5000
  \\
``there is too little talking"
&2000
&2000
  \\
``everyone wants to cook their own soup"
&5000
&2000
  \\
``often a lack of honesty in addressing and dealing with problems in a \\straightforward manner and a sometimes false understanding \\ of political correctness."
&3750
&2420
  \\
``inability to set priorities"
&-99
&2000
  \\
\bottomrule
\end{tabular}
\end{table*}

When an answer is single label, but either BERT or ECC predict multiple labels (82 cases), the true label is almost always among the labels predicted. 
For BERT, the one time the true label was not contained among the predicted labels, the true (hand-coded) label was arguably a mistake: ``social support for German citizens" was hand-coded to label 3700 (``social policy"). BERT instead predicted the codes 2400 (``values, political culture and social criticism") and 3720 (``social justice").
Since the actual single label is almost always included in the multi-label prediction of the models, the question arises whether the additional labels are suitable. We found the additional labels predicted by BERT are always plausible, while ECC predicts some unsuitable labels. 

Some poor ECC predictions when ECC incorrectly predicts multiple labels are listed in Table~\ref{t:eccweird}. 
(Table~\ref{t:appendix2} in the appendix shows the original German text answers.)
For example, for ``fair distribution of taxes" ECC predicted 2400 (``values, political culture, social criticism") and -99 (``no answer"), whereas 4320 (``tax policy") was true.  ECC has several times difficulty with the ``no answer'' label.

\begin{table*}[thbp]
\centering
\caption{Examples of poor multi-label predictions by ECC when the correct code is single label}
\label{t:eccweird}
\begin{tabular}{lrrrrr} \toprule
  & True &  ECC  & ECC  & ECC   \\
  & & Label 1& Label 2 & Label 3\\ \midrule
``fair distribution of taxes"  
& 4320
& 2400
& -99
 \\
 ``small movements are so loud that it makes them seem big" 
&2400
&3412
&3750
  \\
``muslim"

& 3750
& 2411
& -99
 \\
 ``in any case, not the asylum seekers, that is only  &  &  &  \\
   exaggerated" & -99 & 3720 & 3750 \\
``pension or poverty at retirement age"
& 3740
& 3721
& 3740
& 4100
 \\
\bottomrule
\end{tabular}
\end{table*}

Finally, consider answers that are multi-label and ECC or BERT predict multiple labels, but incorrect ones. Both ECC and BERT tend to predict fewer multiple labels than would be correct. 
In extremely rare cases they predict completely wrong labels. One of these rare examples is ``integration of refugees and aging of the German population" for which the true labels are 3750 (``immigration, asylum and integration") and 3740 (``demographic change and pensions") and BERT predicts 3750 (``immigration, asylum and integration") and 2400 (``values, political culture, social criticism"). 

\subsection{An alternative, semi-automatic approach\label{s:semiautomatic}}
Semi-automatic algorithms code some answers automatically, others manually, in an effort to improve  overall accuracy or another metric. 
A semi-automatic alternative to using a pure multi-label strategy is as follows: where BERT predicts a single label, use it. Where BERT predicts multiple or no labels, code manually. 
From  Table~\ref{t:label_distribution} we know that BERT predicts 89.0\% of the text answers to be single label. 
The 0/1 loss of text answers for which BERT predicted 1 label is 0.086. (This result is not in Table~\ref{t:zerooneloss}, which shows loss for the true label distribution).

For these data, the semi-automatic approach suggests to code 11\% of the answers manually. The remainder has a loss of 0.086\%. This approach presents a tradeoff of a reduced loss (8.6\% vs. 13.1\%) and the burden to code 11\% of the answers manually.  Whether the tradeoff is worth it will depend on  the judgement of the analyst. Also,
the strategy may only be realistic in mildly multi-label data sets, as otherwise the fraction of data predicted to be non-single label will increase.

%%%%%%%%%%%%%%%%%%%%%%%%%%%%%%%%%%%%%%%%%%%%%%%%%%%%%%%%%%%%%%%%%%

\section{Discussion\label{s:discussion}}

 Multi-label BERT has several decisive advantages: a) Multi-label BERT convincingly beats the usual multi-label algorithms in terms of 0/1 loss (predicting all labels correctly vs not). 
  Pre-training may be particularly advantageous when the training data are small. Our training data was larger (N=17,584) than the typical training data for open-ended questions in the social sciences, and yet BERT clearly wins. 
b) Because they are hard problems, multi-label approaches have struggled to predict well enough to be useful in settings where high quality is important.  
For example, the 0/1 loss for ECC ranged from  35.8\% to 44.9\% for classifying open-ended questions \citep{schonlau2021all-that-apply}.
For mildly multi-label problems, for the first time, we have reached a 0/1 loss that we believe is low enough to make  (fully) automatic classification attractive. That is, semi-automatic classification is optional rather than necessary. 
c)  Multi-label BERT obtains predictions from a {\em single} model. For the other multi-label algorithms introduced we need one model per label. 
Using multi-label BERT eliminates any need for iterating through the individual models to improve predictions.
d) The analysis of bad predictions showed some of BERT's bad predictions were in fact plausible in light of the coding manual. Further, BERT was much more robust to spelling mistakes than the leading n-gram based approach, ECC.

We see the following limitations: a) Python is required. Social scientists are more familiar with programs like Stata and R. The use of BERT requires programming in Python. We make our code available in the supplementary material to reduce the burden somewhat. 
b) We used German BERT, a language specific model. Since German BERT is less well established than the original BERT,
and because of success for single label classification \citep{gweon2022BERT}, there is no reason to believe this would not work in the English language. 
c) The multi-label algorithms we have shown use SVMs as the base learner. However, we know that 
other algorithms such as gradient boosting and random forest perform similarly when classifying answers to open-ended questions \citep{he2022codingerrors,gweon2022BERT}.

With an average of 1.17 labels per text, the data set we used is mildly multi-label.
What would change if we had a data set with a high average number of labels? 
The 0/1 loss would likely increase because more labels harbour more opportunities for incorrect predictions.
Further, the alternative semi-automatic approach would be infeasible because too many text would have to be coded manually.

For the English language,  \citet{meidinger2021open} found that their baseline model, binary relevance with logistic regression, sometimes beat BERT in terms of 0/1 loss. However, among the 4 data sets that are mildly multi-label (data sets 7-10), BERT and other transformer based models beat the baseline.

Ngram based models tend to work better for short text answers than for longer answers because in short text answers individual words tend to be indicative of the label. Since over half of the answers in our data  were single-word answers, ngram based models have lost to BERT (a transformer based classifier) even on a task that is supposedly their strength. Our result is consistent with recent work on short texts for single label classification  that concludes 
``Transformers are Short Text Classifiers'' \citep{karl2022transformers}, meaning that transformer based text classifiers such as BERT do exceedingly well on short texts.

\section*{Acknowledgement}
We gratefully acknowledge funding from grant 435-2021-0287 from the Canadian Social Sciences and Humanities Research Council (SSHRC)     (PI: Schonlau). We thank Stefan Eschenwecker and Jan-Eric Meurer for their support. 

\section*{Data availability and supplementary material}
Data can be downloaded from \url{https://search.gesis.org/research_data/ZA6838?doi=10.4232/1.13783} \citep{glespanel}. We use only the first wave of the panel. Participants of the first wave can be identified with the variable {\em kp1\_participation}. The text entries of the open ended answers can be found in the separate csv file for wave 1 which can be merged with the other data via the identifier ``lfdn".
The code we used to implement the different models is available as supplementary material.

\FloatBarrier 

\bibliographystyle{chicago}
\bibliography{multilabel}

\begin{thebibliography}{}

\bibitem[\protect\citeauthoryear{Chan, M\"oller, Pietsch, and Soni}{Chan
  et~al.}{2019}]{germanbert}
Chan, B., T.~M\"oller, M.~Pietsch, and T.~Soni (2019).
\newblock German {BERT}.
\newblock \url{https://huggingface.co/bert-base-german-cased}.

\bibitem[\protect\citeauthoryear{Devlin, Chang, Lee, and Toutanova}{Devlin
  et~al.}{2019}]{devlin2019bert}
Devlin, J., M.-W. Chang, K.~Lee, and K.~Toutanova (2019).
\newblock {BERT}: Pre-training of deep bidirectional transformers for language
  understanding.
\newblock arXiv preprint arXiv:1810.04805.

\bibitem[\protect\citeauthoryear{Ford}{Ford}{2023}]{ford2023}
Ford, S.~S. (2023).
\newblock Natural language processing methods for open-ended survey responses:
  The case of the `most important national problem'.
\newblock Talk given on 15 Feb. 2023 at Vanderbilt University and online.

\bibitem[\protect\citeauthoryear{{GLES}}{{GLES}}{2021}]{glespanel}
{GLES} (2021).
\newblock {GLES} {P}anel 2016-2021, {W}ellen 1-15.
\newblock {D}atenarchiv, {K}öln: {ZA}6838. {D}atenfile {V}ersion 5.0.0.
  {DOI}=10.4232/1.13783.

\bibitem[\protect\citeauthoryear{Gweon and Schonlau}{Gweon and
  Schonlau}{2023}]{gweon2022BERT}
Gweon, H. and M.~Schonlau (2023).
\newblock Automated classification for open-ended questions with {BERT}.
\newblock {\em Journal of Survey Statistics and Methodology\/}.
\newblock To appear, arXiv preprint arXiv:2209.06178.

\bibitem[\protect\citeauthoryear{Gweon, Schonlau, and Wenemark}{Gweon
  et~al.}{2020}]{gweon2020semi}
Gweon, H., M.~Schonlau, and M.~Wenemark (2020).
\newblock Semi-automated classification for multi-label open-ended questions.
\newblock {\em Survey Methodology\/}~{\em 46\/}(2), 265--282.

\bibitem[\protect\citeauthoryear{He and Schonlau}{He and
  Schonlau}{2020}]{he2020double_multiple}
He, Z. and M.~Schonlau (2020).
\newblock Automatic coding of open-ended questions into multiple classes:
  Whether and how to use double coded data.
\newblock {\em Survey Research Methods\/}~{\em 14\/}(3), 267--287.

\bibitem[\protect\citeauthoryear{He and Schonlau}{He and
  Schonlau}{2022}]{he2022codingerrors}
He, Z. and M.~Schonlau (2022).
\newblock A model-assisted approach for finding coding errors in manual coding
  of open-ended questions.
\newblock {\em Journal of Survey Statistics and Methodology\/}~{\em 10\/}(2),
  365--376.

\bibitem[\protect\citeauthoryear{Karl and Scherp}{Karl and
  Scherp}{2022}]{karl2022transformers}
Karl, F. and A.~Scherp (2022).
\newblock Transformers are short text classifiers: A study of inductive short
  text classifiers on benchmarks and real-world datasets.
\newblock arXiv preprint arXiv:2211.16878.

\bibitem[\protect\citeauthoryear{Madjarov, Kocev, Gjorgjevikj, and
  D{\v{z}}eroski}{Madjarov et~al.}{2012}]{madjarov2012extensive}
Madjarov, G., D.~Kocev, D.~Gjorgjevikj, and S.~D{\v{z}}eroski (2012).
\newblock An extensive experimental comparison of methods for multi-label
  learning.
\newblock {\em Pattern Recognition\/}~{\em 45\/}(9), 3084--3104.

\bibitem[\protect\citeauthoryear{Meidinger and A{\ss}enmacher}{Meidinger and
  A{\ss}enmacher}{2021}]{meidinger2021open}
Meidinger, M. and M.~A{\ss}enmacher (2021).
\newblock A new benchmark for {NLP} in social sciences: evaluating the
  usefulness of pre-trained language models for classifying open-ended survey
  responses.
\newblock In {\em 13th International Conference on Agents and Artificial
  Intelligence}, pp.\  866--873.

\bibitem[\protect\citeauthoryear{Pedregosa, Varoquaux, Gramfort, Michel,
  Thirion, Grisel, Blondel, Prettenhofer, Weiss, Dubourg, Vanderplas, Passos,
  Cournapeau, Brucher, Perrot, and Duchesnay}{Pedregosa
  et~al.}{2011}]{scikit-learn}
Pedregosa, F., G.~Varoquaux, A.~Gramfort, V.~Michel, B.~Thirion, O.~Grisel,
  M.~Blondel, P.~Prettenhofer, R.~Weiss, V.~Dubourg, J.~Vanderplas, A.~Passos,
  D.~Cournapeau, M.~Brucher, M.~Perrot, and E.~Duchesnay (2011).
\newblock Scikit-learn: Machine learning in {P}ython.
\newblock {\em Journal of Machine Learning Research\/}~{\em 12}, 2825--2830.

\bibitem[\protect\citeauthoryear{Read, Pfahringer, Holmes, and Frank}{Read
  et~al.}{2011}]{read2011classifierchains}
Read, J., B.~Pfahringer, G.~Holmes, and E.~Frank (2011).
\newblock Classifier {C}hains for multi-label classification.
\newblock {\em Machine Learning\/}~{\em 85}, 333--359.

\bibitem[\protect\citeauthoryear{Schierholz and Schonlau}{Schierholz and
  Schonlau}{2021}]{schierholz21occupation}
Schierholz, M. and M.~Schonlau (2021).
\newblock Machine learning for occupation coding — a comparison study.
\newblock {\em Journal of Survey Statistics and Methodology\/}~{\em 9\/}(5),
  1013--1034.

\bibitem[\protect\citeauthoryear{Schonlau and Couper}{Schonlau and
  Couper}{2016}]{schonlau2016semi}
Schonlau, M. and M.~P. Couper (2016).
\newblock Semi-automated categorization of open-ended questions.
\newblock {\em Survey Research Methods\/}~{\em 10\/}(2), 143--152.

\bibitem[\protect\citeauthoryear{Schonlau, Guenther, and Sucholutsky}{Schonlau
  et~al.}{2017}]{schonlau17ngram}
Schonlau, M., N.~Guenther, and I.~Sucholutsky (2017).
\newblock Text mining with n-gram variables.
\newblock {\em Stata Journal\/}~{\em 17\/}(4), 866--881.

\bibitem[\protect\citeauthoryear{Schonlau, Gweon, and Wenemark}{Schonlau
  et~al.}{2021}]{schonlau2021all-that-apply}
Schonlau, M., H.~Gweon, and M.~Wenemark (2021).
\newblock Automatic classification of open-ended questions:
  check-all-that-apply questions.
\newblock {\em Social Science Computer Review\/}~{\em 39\/}(4), 562--572.

\bibitem[\protect\citeauthoryear{Tsoumakas and Katakis}{Tsoumakas and
  Katakis}{2007}]{tsoumakas2007}
Tsoumakas, G. and I.~Katakis (2007).
\newblock Multi-label classification: An overview.
\newblock {\em International Journal of Data Warehousing and Mining
  ({IJDWM})\/}~{\em 3\/}(3), 1--13.

\bibitem[\protect\citeauthoryear{Vaswani, Shazeer, Parmar, Uszkoreit, Jones,
  Gomez, Kaiser, and Polosukhin}{Vaswani et~al.}{2017}]{vaswani2017attention}
Vaswani, A., N.~Shazeer, N.~Parmar, J.~Uszkoreit, L.~Jones, A.~N. Gomez,
  {\L}.~Kaiser, and I.~Polosukhin (2017).
\newblock Attention is all you need.
\newblock {\em Advances in Neural Information Processing Systems
  ({NIPS})\/}~{\em 30}.

\bibitem[\protect\citeauthoryear{Wartena}{Wartena}{2019}]{wartena2019}
Wartena, C. (2019).
\newblock A probabilistic morphology model for {G}erman lemmatization.
\newblock In {\em Proceedings of the 15th Conference on Natural Language
  Processing (KONVENS 2019)}, pp.\  40--49.

\bibitem[\protect\citeauthoryear{Zhang and Zhou}{Zhang and
  Zhou}{2013}]{zhang2013review}
Zhang, M.-L. and Z.-H. Zhou (2013).
\newblock A review on multi-label learning algorithms.
\newblock {\em IEEE Transactions on Knowledge and Data Engineering\/}~{\em
  26\/}(8), 1819--1837.

\end{thebibliography}

\newpage
%%%%%%%%%%%%%%%%%%%%%%%%%%%%%%%%%%%%%%%%%%%%%%%%%%%%%%%%%%%%
\section*{Appendix: Tables with German text answers}
The tables in this appendix contain the original German text answers and have correponding tables in the main paper. 
\begin{table*}[thbp]
\centering
\caption{Original German text answers corresponding to Table~\ref{t:2400}: Incorrectly predicted codes by ECC and BERT when the correct code 2400. Non-text columns are identical to this table.}
\label{t:Appendix1}
\begin{tabular}{lrrrrrr} \toprule
 Answer & ECC &  BERT   \\ \midrule
``sie schauen nicht nach den deutschen"  
& 3100
&   5000
 \\
``egoismus"
 
&    -99
&    5000
 \\
``zufriedenheit schaffen"
&   5000
&   5000

\\
``unklare ziele"
& -99
& 5000
 \\
``bessere erklaerung der politik" 
&2000
&2000
  \\
``zu viele nichtwaehler."
&-99
&-98
  \\
``konsequentes umsetzen"
&-99
&2000
  \\
``einheitlichkeit"
&-99
&5000
  \\
``staendigen verstoße gegen unser grundgesetz und die alternativlose\\ politik von frau merkel"
&2441
&2441
  \\
``kommunikation"
&5000
&5000
  \\
``es wird zu wenig geredet"
&2000
&2000
  \\
``jeder will sein eigenes sueppchen kochen."
&5000
&2000
  \\
``oft fehlende ehrlichkeit, probleme unverkrampft anzusprechen \\und mit diesen umzugehen sowie ein manchmal falsches verstaendnis \\von political correctness"
&3750
&2420
  \\
``unfaehigkeit prioritäten zu setzen"
&-99
&2000
  \\
\bottomrule
\end{tabular}
\end{table*}

\begin{table*}[thbp]
\centering
\caption{Original text answers in German from Table~\ref{t:eccweird}: Examples of poor multi-label predictions by ECC when the correct code is single label. Non-text columns are identical to this table.}
\label{t:appendix2}
\begin{tabular}{lrrrrr} \toprule
  & True &  ECC 1 & ECC 2 & ECC 3   \\ \midrule
``gerechte abgabenverteilung"  
& 4320
& 2400
& -99
 \\
 ``kleine bewegungen so laut sind, dass sie dadurch  & & & \\
 groß wirken" &2400 &3412 &3750 \\
``moslem"
& 3750
& 2411
& -99
 \\
 ``auf jeden fall nicht die asylbewerber, das wird nur  &  &  &  \\
  aufgebauscht" & -99 & 3720 & 3750 \\
``rente bzw.armut im rentenalter"
& 3740
& 3721
& 3740
& 4100
 \\
\bottomrule
\end{tabular}
\end{table*}

\end{document}